\documentclass[twocolumn,floatfix,prl]{revtex4}%
\usepackage[dvipdfmx]{graphicx}%
\usepackage{amsmath}%
\usepackage{amsfonts}%Now
\usepackage{amssymb}
\usepackage{hhline}
\usepackage{bm}
\usepackage{mathrsfs}
\usepackage{color}

\def\s{{\sigma}}
\def\e{{\epsilon}}
\def\k{{ {\bm k} }}
\def\p{{ {\bm p} }}
\def\q{{ {\bm q} }}
\def\Q{{ {\bm Q} }}

\def\0{{ {\bm 0} }}
\def\w{{\omega}}
\def\a{{\alpha}}
\def\b{{\beta}}

\allowdisplaybreaks[4]

\begin{document}
\title{
Prediction of pseudogap formation due to $d$-wave bond-order
in organic superconductor $\kappa$-(BEDT-TTF)$_2$X
}
\author{
Rina Tazai$^1$, Youichi Yamakawa$^1$, Masahisa Tsuchiizu$^2$ and
Hiroshi Kontani$^1$}

\date{\today }

\begin{abstract}

Rich hidden unconventional orders with pseudogap formation,
such as the inter-site bond-order (BO),
attract increasing attention in condensed matter physics.
Here, we investigate the hidden order formation
in organic unconventional superconductor $\kappa$-(BEDT-TTF)$_2$X.
We predict the formation of $d$-wave BO 
at wavelength $\q=\Q_{\rm B}=(\delta,\delta)$ ($\delta=0.38\pi$)
for the first time,
based on both the functional renormalization group (fRG)
and the density-wave equation theories.
The origin of the BO is the quantum interference among 
antiferromagnetic spin fluctuations.
This prediction leads to distinct pseudogap-like reduction
in the NMR $1/T_1$ relaxation rate and in the density-of-states,
consistently with essential experimental reports.
The present theory would be applicable for 
other strongly correlated metals with pseudogap formation.

\end{abstract}

\address{
$^1$Department of Physics, Nagoya University, Furo-cho, Nagoya 464-8602, Japan. \\
$^2$Department of Physics, Nara Women's University, Nara 630-8506, Japan.
}
 
%\keywords{orbital fluctuations, self-consistent vertex correction theory, magnetic quantum criticality}
\sloppy

\maketitle

%%%%%%%%%%%%%%%%%%
%\section{Introduction}
%\label{sec:Intro}
%%%%%%%%%%%%%%%%%
%Organic superconductors are interesting 
%platform of exotic electronic states induced by
%strong Coulomb interaction.

The layered organic compounds 
$\kappa$-(BEDT-TTF)$_2$X
attract considerable attention because of their rich variety of 
ground states due to strong electron correlation.
Many compounds show the antiferromagnetic (AFM) insulating states
at ambient pressure \cite{Kanoda-rev,Kanoda-rev2},
except for several quantum spin-liquid compounds (like X=Cu$_2$(CN)$_3$)
\cite{Shimizu}.
Under pressure, many of them show unconventional superconductivity 
($T_{\rm c}\gtrsim10$K) next to the AFM phase \cite{Kanoda-rev,Kanoda-rev2}.
In X=Cu[N(CN)$_2$]Br and X=Cu(NCS)$_2$,
metallicity and superconductivity appear even at ambient pressure.
Up to now, 
$d_{x^2-y^2}$ and/or $d_{xy}$ symmetries
are predicted based on the spin-fluctuation mechanisms
\cite{Schmalian-ET,Kino-ET,Kondo-ET,Kontani-ET,Kuroki-ET},
%the variational Monte Carlo
the VMC study
\cite{Watanabe},
%cluster dynamical mean-field
and cluster DMFT study
\cite{Tremblay}.
%cluster perturbation 
%\cite{CPT}
%theoies.

A central mystery in anomalous metallic states 
in $\kappa$-(BEDT-TTF)$_2$X would be the origin of the pseudogap 
and its relation to the superconductivity.
Both the NMR relaxation ratio $1/T_1T$ 
\cite{Kanoda-rev,Kanoda-rev2}
and the density-of-states (DOS) measured by the STM
\cite{Nomura-STM}
exhibit gap behaviors below $T^*\sim 50$K.
As origins of the pseudogap, for example,
crossover scenarios due to
strong spin or superconducting (SC) fluctuations
\cite{Kawamoto1,Kawamoto2,Kontani-rev,Yamada-rev,Moriya-rev,Schmalian-PG,Tremblay-PG,Tsuchiya1,Tsuchiya2}
and proximity effect to Mottness
\cite{Mottness-PG,Kang}
have been discussed.
However, distinct kink-like changes observed in $1/T_1T$ 
\cite{Kanoda-rev,Kanoda-rev2}
and in the hardening of optical phonon frequency
\cite{Raman}
indicates the emergence of a hidden order parameter
at $T\approx T^*$.
(The intra-dimer charge disproportionation is reported
only in X=Cu$_2$(CN)$_3$ \cite{Watanabe}.)

The similarity between
$\kappa$-(BEDT-TTF)$_2$X and cuprate high-$T_{\rm c}$ superconductors 
has been actively discussed for years.
Recently, in many cuprates, the charge density-wave (DW) order 
in period $3a\sim4a$ at $T_{\rm CDW}\sim200$K
has been discovered by 
the X-ray scattering, STM study, and NMR analysis
\cite{Y-Xray,Bi-Xray,STM-Kohsaka,STM-Fujita,NMR-Julien1,NMR-Julien2}.
In addition, nematic transition 
(presumably at $\q={\bm 0}$) emerges at the
pseudogap temperature $T^* \ (>T_{\rm CDW})$ 
\cite{Y-Sato,Hg-Murayama,Fujimori-nematic,Shibauchi-nematic,Zheng-NMR}.
These charge-channel phase transitions at $T_{\rm CDW}$ and $T^*$
have been hotly discussed based on the 
charge/spin current order
\cite{Varma,Yokoyama,Kontani-sLC}, 
and $d$-wave bond-order (BO) scenarios
\cite{Bulut,Chubukov,Chubukov-AL,Sachdev,Holder,DHLee-PNAS,Kivelson-NJP,Yamakawa-CDW,Tsuchiizu-CDW1,Tsuchiizu-CDW2,Kawaguchi-CDW}. 
The BO is the modulation of correlated hopping integrals,
and it can be 
caused by the paramagnon interference mechanism 
\cite{Yamakawa-CDW,Tsuchiizu-CDW1,Tsuchiizu-CDW2,Kawaguchi-CDW}
that is overlooked in usual spin fluctuation theories.
This mechanism has also been applied to Fe-based superconductors
\cite{Onari-SCVC,Onari-FeSe,Yamakawa-FeSe,Onari-B2g,Onari-AFBO}
and heavy-fermion systems
\cite{Tazai-CeB6}.
Thus, it is significant to study the 
paramagnon interference effects in $\kappa$-(BEDT-TTF)$_2$X
to go beyond the conventional understanding of this system.
%to understand its anomalous metallic states.

%punch line
In this paper,
we discuss the occurrence of hidden charge DW orders
in $\kappa$-(BEDT-TTF)$_2$X
% in an unbiased way,
by using both the functional renormalization group (fRG) 
and the DW equation methods.
We predict the $d$-wave BO order formation 
at wavelength $\q=\Q_{\rm B}=(\delta,\delta)$ ($\delta=0.38\pi$)
in $\kappa$-(BEDT-TTF)$_2$X for the first time.
The origin of the BO is novel quantum interference among 
paramagnons with $\Q_{\rm S}^{\pm}\approx(\pi\pm\delta/2,\pi\pm\delta/2)$.
The predicted distinct reductions in the spin fluctuation strength
and in the DOS at BO transition temperature 
naturally explain the NMR and Raman measurements.

%%%%%%%%%%%%%%%%%%%%%%%%%%%%%%%%%%%%%%%
\begin{figure}[htb]
\includegraphics[width=.85\linewidth]{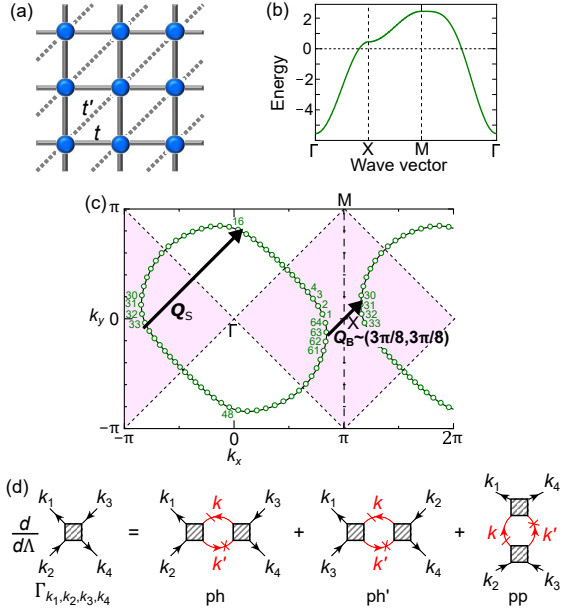}
\caption{
(a) Anisotropic triangular lattice 
for $\kappa$-(BEDT-TTF)$_2$X.
(b) Band dispersion and (c) the FS at $n=1$.
The major nesting vector is $\Q_{\rm S}\approx(\pi-\delta/2,\pi-\delta/2)$
and the minor nesting vector is $\Q_{\rm B}\approx(\delta,\delta)$
with $\delta\approx3\pi/8$.
(d) Differential equation of the fRG theory.
The crossed line represents electron propagator 
inside of the energy cutoff ($|\e_{\k'}|<\Lambda_{l}$), 
while the slashed line denotes on-shell one ($|\e_{\k}|=\Lambda_{l}$).
}
\label{fig:model}
\end{figure}
%%%%%%%%%%%%%%%%%%%%%%%%%%%%%%%%%%%%%%%

To analyze the 
quantum interference in low-dimensional metals, the fRG method
is very powerful and reliable, since various DW instabilities
(both particle-particle and particle-hole channels)
are treated on the same footing
%on the same footings
\cite{Holder,Metzner-rev,Honerkamp,Husemann,Tsuchiizu-2013,Tsuchiizu-CDW1,Tsuchiizu-CDW2,Tazai-FRG}.
Using the fRG, unconventional DW states in cuprates
and ruthenates have been analyzed satisfactorily
\cite{Tsuchiizu-2013,Tsuchiizu-CDW2}.
%the $d$-wave bond-order (BO),
%which is the antiphase modulation of correlated hopping integrals,
%originatesd from the paramagnon interference
%in cuprate superconductors
%at wavevectors $\q \approx (\pi/2,\pi/2)$ and $\q=\bm{0}$
%This result naturally explains the phase diagram in cuprates. 
The fRG method is suitable to analyze the many-body electronic states,
%without any assumption, 
and we can obtain reliable results by making careful comparison 
with the diagrammatic calculation 
using the DW equation method
\cite{Onari-FeSe,Tsuchiizu-CDW2,Onari-B2g,Onari-AFBO}.

%%%%%%%%%%%%%%%%%%
%\section{Model and fRG study}
%\label{sec:fRG}
%%%%%%%%%%%%%%%%%%

Here, we introduce the anisotropic triangular lattice 
dimer Hubbard model, which is the simplest effective model 
for $\kappa$-(BEDT-TTF)$_2$X; $\hat{H}=\hat{H}_{0}+\hat{H}_{U}$
\cite{Kino-Fukuyama}.
The kinetic term is given by 
$\hat{H}_{0}=\sum_{\k\sigma}\epsilon_{\k}c^{\dagger}_{\k\sigma}c_{\k\sigma}$
with $\epsilon_{\k}=2t(\cos k_x+\cos k_y)+2t'\cos(k_x+k_y)$.
Indices $\k$ and $\sigma$ denote the momentum and spin, respectively.
Here, we set the hopping integrals in Fig. \ref{fig:model} (a)
as $(t,t')=(-1,-0.5)$.
We verified that similar numerical results are obtained for 
$t'/t=0.5\sim 0.8$,
which is realized in many compounds
\cite{McKenzie}.
%The tight binding model in the real space is shown in Fig.\ref{fig:model} (a). 
%Enrgy scale is mesuared from Fermi energy.
The on-site Coulomb interaction term is given by
$\hat{H}_{U}=\sum_{k_1,k_2,k_3,k_4}
\frac14 \Gamma^0_{k_1,k_2,k_3,k_4} c^{\dagger}_{k_1}c_{k_2}c_{k_3}c^{\dagger}_{k_4}
\cdot \delta_{\k_1-\k_2-\k_3+\k_4,\bm{0}}$,
where $k_i=(\k_i,\s_i)$
and $\bm{0}$ is the zero vector in the reduced zone.
The initial four-point vertex is
$\Gamma^0_{k_1,k_2,k_3,k_4}
=U(\delta_{\s_1,\s_3}\delta_{\s_2,\s_4}-\delta_{\s_1,\s_2}\delta_{\s_3,\s_4})$.
%where $c^{\dagger}$ ($c$) is a creation 
%(annihilation) operator. Index $\k$, $i$ and $\sigma$ denote momentum, site and spin, respectively.
%$\Gamma^{\sigma \sigma' \rho \rho'}$ includes intra-orbital $U$. 
The four-point vertex 
%$\Gamma^{\sigma \sigma' \rho \rho'}_{\k+\q,\k;\k',\k'-\q}$
acquires the momentum dependence after the renormalization,
which is uniquely divided into the spin and charge parts as
$\Gamma_{k_1,k_2,k_3,k_4}^0= 
\frac12 \Gamma_{\k_1,\k_2,\k_3,\k_4}^{0c}\delta_{\s_1,\s_2}\delta_{\s_4,\s_3}
+
\frac12 \Gamma_{\k_1,\k_2,\k_3,\k_4}^{0s}(\vec{\s})_{\s_1,\s_2}\cdot(\vec{\s})_{\s_4,\s_3}$.
The initial condition
$\Gamma^{0s}=-\Gamma^{0c}\ (=U)$ is largely 
modified by the renormalization.

In the following numerical study,
we set the energy unit $|t|=1$, 
and put the temperature $T=0.05$
and the electron filling $n=1$ ($\mu=0.55$).
The band structure and the Fermi surface (FS)
are presented in Figs. \ref{fig:model} (b) and (c), respectively.
The patch indices ($1\sim64$) are shown on the ellipsoid electron pockets.
The total band width is $W_{D}\sim 10$ (in unit $|t|=1$), and
$|t|$ corresponds to 0.05eV since $W_{D}\sim 0.5$eV experimentally 
\cite{Kino-Fukuyama,Kino-ET}.

In the present study, 
we analyze the model by applying the RG+cRPA method
developed in Refs. 
\cite{Tsuchiizu-2013,Tsuchiizu-CDW1,Tsuchiizu-CDW2,Tazai-FRG},
which is an efficient hybrid method between the fRG and the 
random-phase-approximation (RPA).
Here, we introduce the higher-energy cutoff $\Lambda_0 \ (=2)$
and the logarithmic energy mesh $\Lambda_{l}=\Lambda_{0}e^{-l}$ with $l\ge0$.
Then, the effective four-point vertex $\hat{\Gamma}$ is 
derived from the one-loop RG equation in Fig. \ref{fig:model} (d),
by including the on-shell contribution
$\Lambda_{l+dl}<|\e_\k-\mu|\le \Lambda_l$ step-by-step.
In strongly correlated systems,
$\hat{\Gamma}$ strongly deviates from $\hat{\Gamma}^0$.
%Then, $\hat{\Gamma}$ depends on $l$
%inside the energy region $|\e_\k-\mu|\le \Lambda_0$,
%by applying the logarithmic energy mesh
%$\Lambda_{l}=\Lambda_{0}e^{-l}$ with $l\ge0$, and
In this procedure,
the Brillouin zone (BZ) is divided into $N_p$ patches; 
see the Supplemental Materials (SM) A \cite{SM}.
Below, we set $\Lambda_0 =2$ and $N_p=64$.
(Numerical results are not sensitive 
to the choices of $\Lambda_0$ and $N_p$.)
We analyze the higher energy region $|\e_\k-\mu|\ge \Lambda_0$,
where the vertex corrections (VCs) are less important,
based on the constrained RPA (cRPA) 
with high numerical accuracy using fine $\k$-meshes.
By using the RG+cRPA method,
we can perform reliable calculations
%thanks to the combination of the cRPA and the fRG formalism 
\cite{Tsuchiizu-2013,Tsuchiizu-CDW1,Tsuchiizu-CDW2}.

\begin{figure}[htb]
\includegraphics[width=.9\linewidth]{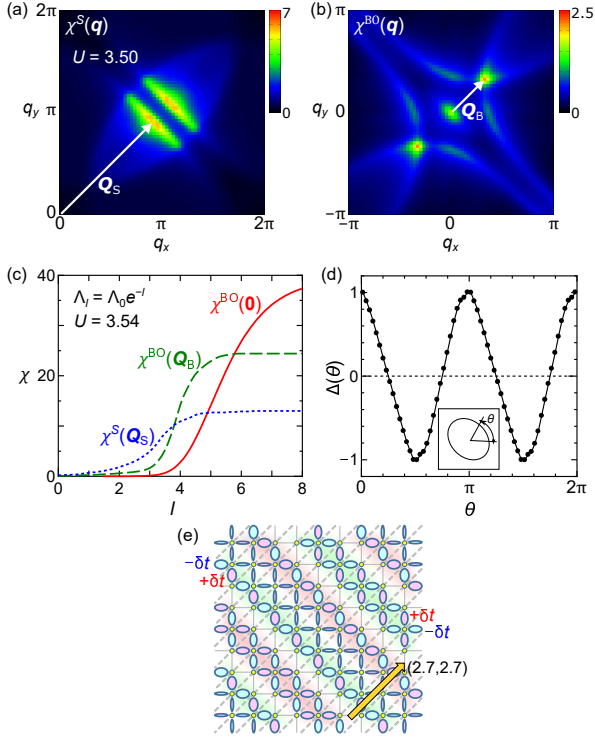}
\caption{
The $\q$-dependences of (a) $\chi^{\rm S}(\q)$ and (b) $\chi^{\rm BO}(\q)$
obtained by the RG+cRPA method at $U=3.5$.
(c) The RG flow for spin and BO susceptibilities at $U=3.54$.
(d) Obtained optimized SC gap function, which 
belongs to $d_{x^2-y^2}$-wave symmetry.
(e) Schematic BO pattern at $\q=\Q_{\rm B}$ in the real space,
where $\bm{\lambda}\approx(8/3,8/3)$ is the wavevector.
%by setting $\Q_{\rm B}=$
%${\bm\lambda}=(2.7a,2.7a)$,
}
\label{fig:bond1}
\end{figure}

In Fig. \ref{fig:model} (d),
the 3rd (1st and 2nd) term of r.h.s represents scattering process 
for particle-particle (pp) channel
(particle-hole (ph) channels).
%Here, we adopt 1-loop approximation and neglect higher than 6-point vertex.
The RG flow starts from $l=0$ ($\Lambda_{l}=\Lambda_0$) to 
$l\rightarrow\infty$ ($\Lambda_{l}\rightarrow0$).
In the parallel way, we solve the RG equation for the susceptibilities
\cite{Tsuchiizu-2013}.
The RG flow will stop for 
$\Lambda_l\lesssim \w_c$ with $\w_c={\rm max}\{ T,\gamma\}$,
where $\gamma \ (\propto|{\rm Im}\Sigma|)$ is the quasiparticle damping rate.
Considering large $\gamma$ in $\kappa$-(BEDT-TTF)$_2$X,
we introduce the low-energy cutoff
$\w_c=\pi T$ in the RG equation of the four-point vertex $\Gamma$
in calculating Fig. \ref{fig:bond1}
by following Refs. \cite{Tsuchiizu-CDW1,Tsuchiizu-CDW2}.

%%%%%%%%%%%%%%%%%%%%%%%%%%%%%%%%
%\section{Bond-ordered phase and superconductivity}
%\label{sec:suscep}
%%%%%%%%%%%%%%%%%%%%%%%%%%%%%%%%

Here, we show numerical results for the dimer Hubbard model.
%In the following calculation, we use $512\times 512$ $\k$-meshes 
%and $1024$ Matsubara frequencies.
First, we calculate various DW susceptibilities at $\w=0$
using the RG+cRPA method.
The spin (charge) susceptibility with the form factor $f_x(\k)$ is
% given as
\cite{Tsuchiizu-CDW2}
%which is obtained from 
%the 3-point vertex function by cRPA+fRG \cite{Tsuchiizu4}.
%The definition of bond (spin) suseptibility; $\chi^{BO(S)}(q)$ is written as
%
\begin{eqnarray}
\chi^{s(c)}_x(\q)=\int^{\beta}_{0} d\tau \frac{1}{2}\left\langle A^{s(c)}_x({\bm q},\tau)
A^{s(c)}_x({\bm -\q},0)\right\rangle, 
\end{eqnarray}
%
%where $q\equiv (\q,\omega_n)$. $\omega_n=2n\pi T$ is Boson Matsubara frequency.
where
$A^{s(c)}_x({\bm q})\equiv \sum_{\k}
f_x(\k) (c^{\dagger}_{{\k-\q/2}  \uparrow}c_{{\k+\q/2} \uparrow}
-(+)c^{\dagger}_{{\k-\q/2} \downarrow}c_{{\k+\q/2} \downarrow})$.
%The spin susceptibiity is given as 
We calculate the spin and charge susceptibilities
with the following form factors $f_1=1$, 
$f_x=\sqrt{2}\sin k_x$, $f_y=\sqrt{2}\sin k_y$, 
$f_{x^2-y^2}=\cos k_x - \cos k_y$ and $f_{xy}=2\sin k_x \sin k_y$.
As a result, 
we find that the spin susceptibility $\chi^{\rm S}(\q) \equiv \chi^s_1(\q)$
and the $d_{x^2-y^2}$-wave BO susceptibility
$\chi^{\rm BO}(\q)\equiv \chi^c_{x^2-y^2}(\q)$ strongly develop.
Other susceptibilities remain small in the present study.

%%%%%%%%%%%%%%%%%%%%%%%%%%%%%%%%%%%%%%%%%%%%%
%%%%%%%%%%%%%%%%%%%%%%%%%%%%%%%%%%%%%%%%%%%%%

In Figs. \ref{fig:bond1} (a) and (b), we plot $\q$-dependences of 
 $\chi^{\rm S}(\q)$ and $\chi^{\rm BO}(\q)$ at $U=3.5$.
%in the case of $\w_c=\pi T$.
Strong spin fluctuations develop at $\q=\Q_S$,
consistently with the previous RPA and FLEX analyses \cite{Kino-ET}. 
In addition, we reveal the development of 
$\chi^{\rm BO}(\q)$ at $\q=\Q_{\rm B}\approx (3\pi/8,3\pi/8)$
in addition to $\q=(0,0)$.
%Since $\chi^{\rm BO}(\q)$ is not enhanced at all in the RPA and FLEX,
The obtained strong bond fluctuations originate from the VCs
that are dropped in the RPA.
%VCs that are dropped in the RPA (FLEX).
%as we will discuss in detail later.

The $\chi^{\rm BO}(\q)$ strongly develops by increasing $U$.
Figure \ref{fig:bond1} (c) shows the RG flow of the 
susceptibilities in the case of $U=3.54$.
In this case, the bond susceptibility exceeds the spin one
after completing the renormalization.
We see that $\chi^{\rm S}(\Q_S)$ starts to increase
in the early stage of the renormalization,
by reflecting the major nesting of the FS at $\q=\Q_S$.
Next, $\chi^{\rm BO}(\Q_{\rm B})$ starts to increase for $l\gtrsim3$,
and it exceeds $\chi^{\rm S}(\Q_S)$ at $l\sim4$.
Finally, 
%the ferro-BO susceptibility 
$\chi^{\rm BO}(\bm{0})$
starts to increase for $l\gtrsim 4$ ($\Lambda_l\lesssim 0.037$),
because the renormalization of the Pauli ($\q=\bm{0}$) susceptibility
occurs only for $\Lambda_l\lesssim T$.
All susceptibilities saturate for $l\gtrsim8$ 
($\Lambda_l\lesssim 0.7\times10^{-3}$).
The final results in Figs. \ref{fig:bond1} (a) and (b)
are given at $l\approx9$.
Thus, all $\chi^{\rm S}(\Q_S)$, $\chi^{\rm BO}(\Q_{\rm B})$ and 
$\chi^{\rm BO}(\bm{0})$ strongly develop at $U=3.54$.
%The final susceptibilities are given by $l\sim 9$
%($\Lambda_l\lesssim T/100$) in the RG flow.

\begin{figure}[htb]
\includegraphics[width=.9\linewidth]{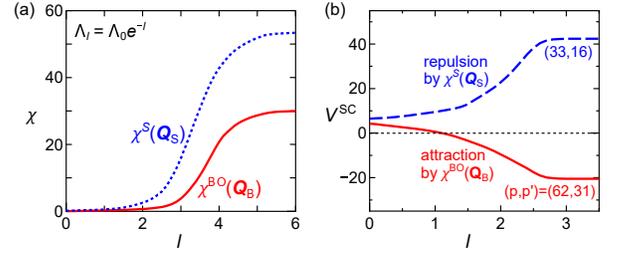}
\caption{
Obtained RG results in the case of $\omega_c^{\rm pp}=10T$ at $U=3.82$.
(a) Obtained RG flow for $\chi^{\rm BO}(\Q_{\rm B})$ and $\chi^{\rm S}(\Q_S)$.
(b) Obtained pairing interaction $V^{\rm SC}(\p,\p')$
due to the spin-fluctuation-mediated repulsion
(for $(\p,\p')=(33,16)$)
and the BO fluctuation-mediated attraction
(for $(\p,\p')=(62,31)$).
The patch number $\p$ and $\p'$ are shown in Fig. \ref{fig:model} (c).
}
\label{fig:10piT}
\end{figure}

We also calculate the spin-singlet SC susceptibility
\cite{Tsuchiizu-CDW2}
%, which is defined as
%
\begin{eqnarray}
\chi^{\rm{SC}}= \frac{1}{2}\int^{\beta}_{0} d\tau
\left\langle B^{\dagger}(\tau)B(0)\right\rangle,
\left( B\equiv \sum_{{\bm k}}\Delta({\bm k})
c_{{\bm k}\uparrow}c_{{\bm -\k}\downarrow} \right)
\label{eqn:B}
\end{eqnarray}
where $\Delta({\bm k})$ is an even parity gap function,
which is uniquely determined so as to maximize $\chi^{\rm{SC}}$
under the constraint 
$\frac1N \sum_\k|\Delta({\bm k})|^2\delta(\e_\k-\mu) =1$
\cite{Tsuchiizu-CDW2}.
We show the obtained optimized gap at $U=3.54$ in Fig. \ref{fig:bond1} (d).
The obtained gap function in the $d_{x^2-y^2}$-wave symmetry 
is understood as the spin-fluctuation-mediated $d$-wave state
\cite{Kino-ET}.
In the present case, large BO susceptibilities $\chi^{\rm BO}(\Q_{\rm B})$
and $\chi^{\rm BO}(\bm{0})$ should contribute to the pairing mechanism
(see Fig. \ref{fig:10piT} (b)).

\begin{figure}[htb]
\includegraphics[width=.9\linewidth]{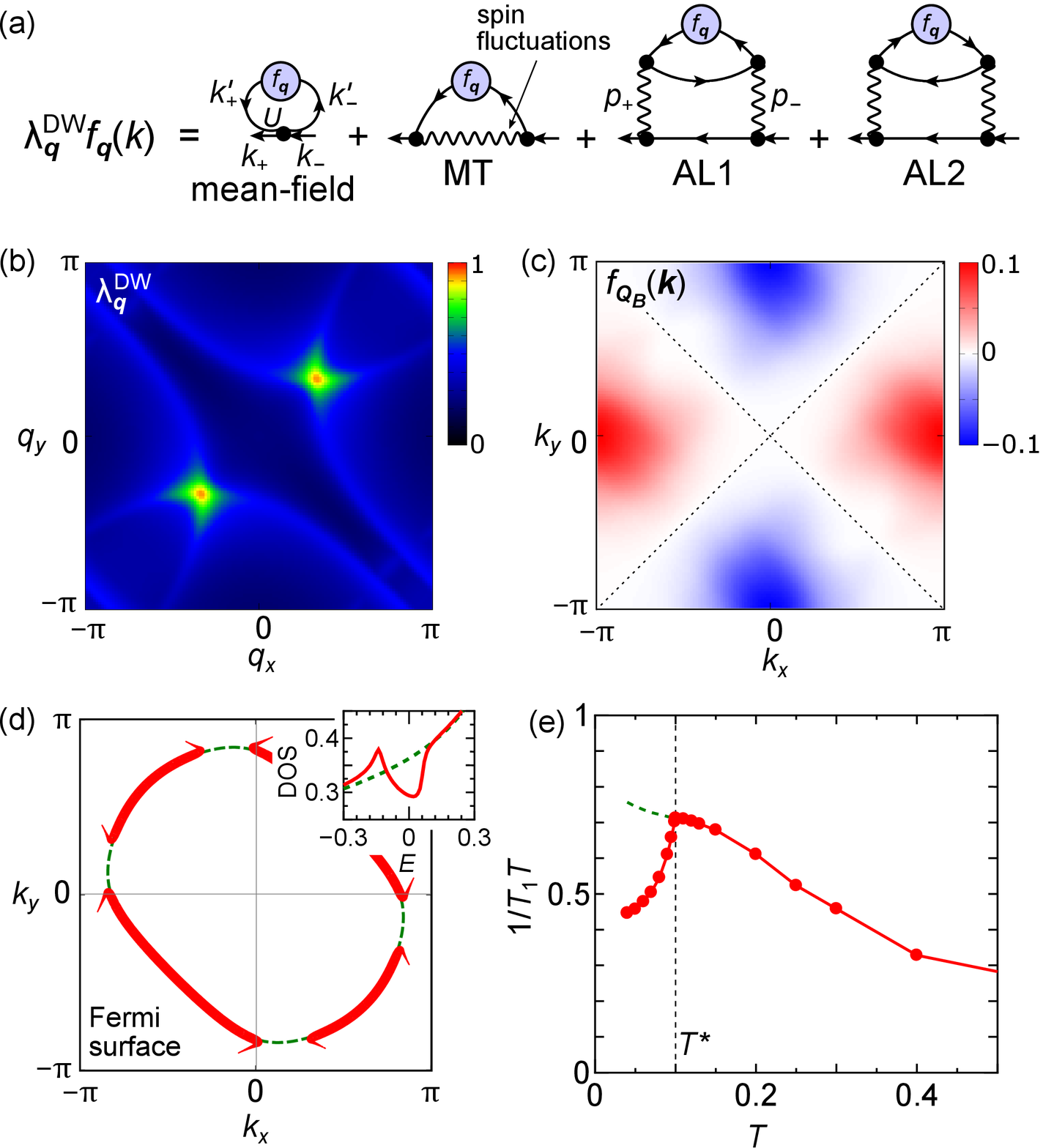}
\caption{
(a) The DW equation for wavevector $\q$.
$\k_\pm \equiv \k\pm\q/2$ and $\p_\pm \equiv \p\pm\q/2$.
The kernel function is composed of the mean-field, MT, and
AL1,2 terms. 
%The AL1 (AL2) term contains the ph (pp) channel.
These diagrams are also produced systematically by solving the RG equation.
(b) Obtained eigenvalue of the BO $\lambda^{\rm DW}_\q$ and 
(c) the form factor $f_\q(\k)$ at $\q=\Q_{\rm B}$ and $\e_n=\pi T$.
(d) Obtained Fermi arc structure in the unfolded zone,
and the pseudogap in the DOS with $f^{\rm max}=0.1$ is shown in the inset.
%(e) the pseudogap in the DOS caused by the finite BO 
(e) Obtained $1/T_1T$, where $T^*$ is the BO transition temperature.
}
\label{fig:fig4}
\end{figure}

Figure \ref{fig:bond1} (e)
shows the schematic $d$-wave BO pattern at $\q=\Q_{\rm B}$.
Here, each red (blue) ellipse represents the increment (decrement)
of the hopping integral $\delta t_{\mu}$ ($\mu=x,y$)
caused by the BO parameters.
The opposite sign between the adjacent $\delta t_{x}$ and $\delta t_{y}$ 
reflects the $d$-wave symmetry of the BO.
The BO parameter causes the pseudogap in the DOS
(see Figs. \ref{fig:fig4} (d) and (e)).
%as we will explain later.

%%%%%%%%%%%%%%%%%%%%%%%%%%%%%%%%
%%%%%%%%%%%%%%%%%%%%%%%%%%%%%%%%

Now, we clarify the importance of the spin fluctuations
on the BO fluctuations.
For this purpose, we solve the RG equation by dropping the 
contribution from the pp channel in the RG equation 
for $\Gamma$,
by introducing an additional cutoff energy 
only for the pp channel; $\omega_c^{\rm pp} (>\w_c)$.
%Then, the contribution from 
%the 3rd term in the r.h.s of Fig.\ref{fig:model} (d)
%is dropped for $\Lambda_l<\w^{\rm pp}$.
Here, we set $\omega_c^{\rm pp}=10T$ to suppress
the SC fluctuations selectively.
%while $\s^S(\Q_S$ strongly develops thanks to the Peierls channels.
The obtained RG flows of the susceptibilities 
at $U=3.82$ are shown in Fig. \ref{fig:10piT} (a).
We see that $\chi^{\rm S}(\Q_S)$ starts to increase in the early stage,
due to the ph channels in the RG equations in 
Fig. \ref{fig:model} (d).
Next, $\chi^{\rm BO}(\Q_{\rm B})$ also increases 
to follow the increment of $\chi^{\rm S}(\Q_S)$, 
similarly to Fig. \ref{fig:bond1} (c).
This result strongly indicates that the BO fluctuations
are driven by the spin fluctuations.
Note that
$\chi^{\rm BO}(\Q_{\rm B})$ exceeds $\chi^{\rm S}(\Q_S)$
by setting $U=3.86$
even in the case $\w_c^{\rm pp}=10T$.

Next, we discuss the SC pairing vertex function
$V^{\rm SC}(\p,\p')=\frac32 \Gamma^s(\p,\p',-\p',-\p)
-\frac12 \Gamma^c(\p,\p',-\p',-\p)-\Gamma^{0s}$
given by the RG+cRPA method
\cite{Tazai-FRG}.
Due to large $\omega_c^{\rm pp} \ (=10T)$,
the obtained $V^{\rm SC}(\p,\p')$ becomes 
``irreducible with respect to the pp channel'' below $\w_c^{\rm pp}$.
Then, $V^{\rm SC}(\p,\p')$ gives the 
``pairing interaction in the SC gap equation''
with the BCS cutoff energy $\w_{\rm BCS}=\omega_c^{\rm pp}$.
The obtained RG flow of $V^{\rm SC}(\p,\p')$ is shown in 
Fig. \ref{fig:10piT} (b).
The large repulsion for $\p-\p'\approx\Q_S$
is apparently given by the spin fluctuations.
Interestingly, we find that the attraction for $\p-\p'\approx\Q_{\rm B}$
is caused by the BO fluctuations in Fig. \ref{fig:10piT} (a).
The present result indicates that 
both $\chi^{\rm BO}(\Q_{\rm B})$ and $\chi^{\rm S}(\Q_S)$ 
cooperatively work as the 
pairing glue of the $d_{x^2-y^2}$-wave state,
as understood in Fig. \ref{fig:model} (c).
The obtained gap structure is very similar to 
Fig. \ref{fig:bond1} (d).
Therefore, the $d$-wave BO fluctuations in the single-orbital 
Hubbard model can mediate large attractive pairing interaction.

Next, we explain that the BO fluctuations
originate from the quantum interference between paramagnons,
which is described by the Aslamazov-Larkin (AL) quantum process.
For this purpose, we analyze the following DW equation 
\cite{Onari-FeSe,Kawaguchi-CDW,Onari-B2g,Onari-AFBO}:
\begin{eqnarray}
\lambda_{\q}^{\rm DW}f_\q(k)= 
-\frac{T}{N}\sum_{k'}I_\q^c(k,k')G(k'_-)G(k'_+)f_\q(k') ,
\label{eqn:DW} 
\end{eqnarray}
where $\lambda_{\q}^{\rm DW}$ is the eigenvalue 
that represents the charge channel DW instability
at wavevector $\q$.
Here, $\bm{p}_\pm \equiv \p\pm\q/2$, and
$k\equiv (\k,\e_n)$ and $p\equiv (\p,\e_m)$
($\e_n$, $\e_m$ are fermion Matsubara frequencies).
The eigenfunction $f_\q(k)$ gives the form factor.
The corresponding DW susceptibility is 
$\chi^{\rm BO} \propto(1-\lambda_\q^{\rm DW})^{-1}$.
In the present model, the obtained DW state
corresponds to the $d$-wave BO.

%Equation (\ref{eqn:DW}) is interpreted as the 
%`` electron-hole pairing equations''. 
The kernel function $I^c$
is given by the Ward identity $-\delta\Sigma/\delta G$,
which is composed of 
one single-paramagnon exchange term and two double-paramagnon exchange ones:
The former and the latter are called 
the Maki-Thompson (MT) term and the AL terms; 
see Fig. \ref{fig:fig4} (a).
(Each wavy line is proportional to $\chi^{\rm S}(\q)$.)
%The lowest order Hartree term $-U$
%in $I_\q^{c}$ gives the RPA contribution.
Among four terms in $I_\q^{c}$,
the AL terms are significant for $\a_S\lesssim1$,
and they give spin-fluctuation-driven
nematic orders in cuprates and Fe-based superconductors
\cite{Onari-SCVC,Kawaguchi-CDW}.

%%%%%%%%%%%%%%%%%%%%%%%%%%%%%%%%
%%%%%%%%%%%%%%%%%%%%%%%%%%%%%%%%

Figure \ref{fig:fig4} (b) shows the obtained 
eigenvalue of the BO, $\lambda_{\q}^{\rm DW}$,
at $T=0.05$ and $U=2.53$,
where the Stoner factor $\a_S=U\chi^0(\Q_S)$ is $0.90$.
Here, $\chi^0(\q)$ is the irreducible susceptibility,
and the SDW occurs when $\a_S=1$.
The eigenvalue $\lambda_{\q}^{\rm DW}$ 
reaches almost unity at $\q=\Q_{\rm B}$,
which is consistent with $\chi^{\rm BO}(\q)$ given by the 
RG+cRPA in Fig. \ref{fig:bond1} (b).
The corresponding form factor in Fig. \ref{fig:fig4} (c)
possesses the $d_{x^2-y^2}$-wave symmetry.
%(Note that $\lambda_{\q}^{\rm BO}$ shows a peak structure at $\q={\bm 0}$
%for $\a_S\sim0.99$ for $t'/t=0.7$; see the SM B \cite{SM}.)
Therefore, strong $d_{x^2-y^2}$-wave BO susceptibility 
at $\q=\Q_{\rm B}$ by the RG+cRPA method in Fig. \ref{fig:bond1} (b)
is well reproduced by the DW equation.
The origin of the strong BO instability at $\q=\Q_{\rm B}$
is the AL terms in Fig. \ref{fig:fig4} (a)
that represent the quantum interference among paramagnons
at $\Q_{\rm S}^{\pm}\approx(\pi,\pi)\pm\Q_{\rm B}/2$.
(Note that $\chi^{\rm S}(\q)$ given by the RPA is similar to 
that the fRG result in Fig. \ref{fig:bond1} (a).)

The paramagnon-interference mechanism can generate
both the ferro-BO instability
(at $\q=\Q_{\rm S}^{\pm}-\Q_{\rm S}^{\pm}=\bm{0}$)
and the incommensurate-BO one
(at $\q=\Q_{\rm S}^{+}-\Q_{\rm S}^{-}=\Q_{\rm B}$).
This mechanism causes the ferro-BO states
in both Fe-based and cuprate superconductors
according to the DW equation analysis
\cite{Onari-FeSe,Onari-B2g,Kawaguchi-CDW,Onari-AFBO}.
%meaning that the ferro-BO fluctuations originates solely 
%from the spin fluctuations.
In the present dimer Hubbard model, in contrast,
the ferro-BO fluctuations remain small
in the DW equation analysis.
This is also true in the fRG analysis
with $\omega_c^{\rm pp}=10T$ shown in Fig. \ref{fig:10piT}.
These results indicate that the 
paramagnon interference mechanism alone
is not sufficient to establish large $\chi^{\rm BO}(\bm{0})$
in Fig. \ref{fig:bond1} (b).
Therefore, we conclude that large $\chi^{\rm BO}(\bm{0})$ 
%in Fig. \ref{fig:bond1} (b)
is caused by the spin and SC fluctuations cooperatively,
since the AL processes by SC fluctuations can cause 
the ferro-BO fluctuations according to 
Ref. \cite{Tsuchiizu-2013}.

Finally, we discuss the band-folding and hybridization gap 
due to the BO with $\q=\Q_{\rm B}$.
Figure \ref{fig:fig4} (d) shows the Fermi arc structure
obtained for $f^{\rm max}\equiv \max_\k \{f_{\Q_{\rm B}}(\k)\} =0.1$.
Here, the folded band structure under the BO at $\q=\Q_{\rm B}$ 
is ``unfolded'' into the original Brillouin zone \cite{Ku}
to make a comparison with ARPES experiment.
The resultant pseudogap in the DOS is shown in 
the inset of Fig. \ref{fig:fig4} (d),
which is consistent with the STM study
\cite{Nomura-STM}.
%The unfolded band structure in the single-$\Q_{\rm d}$ sLC order
%is displayed in Fig. \ref{fig:figS2} in the SM B \cite{SM}.
The BO leads to significant reduction 
of the spin fluctuation strength,
so the obtained $1/T_1T \propto \sum_{\q,\a,\b} 
{\rm Im}\left. \chi^s_{\a,\b}(\q,\w)/\w \right|_{\w=0}$
shown in Fig. \ref{fig:fig4} (e)
exhibit kink-like pseudogap behavior.
Here, $\a,\b$ represent the sites in the unit cell under the presence of the BO,
and we set $f^{\rm max}=0.2\times{\rm tanh}(1.74\sqrt{(1-T/T^*)}$ 
below the BO transition temperature $T^*=0.1$.
(Here, $2f^{\rm max}(T=0)/T^*=4$.)
The obtained pseudogap behaviors in $1/T_1T$ and DOS 
%due to the bond-order formation 
are consistent with phase-transition-like experimental behaviors
\cite{Kanoda-rev,Kanoda-rev2,Raman}.

%add
In the fRG study, the parquet VCs are generated by 
considering all ph, ph' and pp channels in Fig. \ref{fig:model} (d).
On the other hand, in the DW equation study, 
the VCs are limited to MT and AL terms,
whereas their frequency dependences are calculated correctly.
Both theoretical methods lead to the 
emergence of the {\it same} $d$-wave bond order shown in 
Fig. \ref{fig:bond1} (e).

In summary,
we predicted the emergence of the $d$-wave BO 
at wavevector $\q=\Q_{\rm B}=(0.38\pi,0.38\pi)$
in $\kappa$-(BEDT-TTF)$_2$X,
due to the interference between paramagnons
with $\Q_{\rm S}^{\pm}\approx(\pi,\pi)\pm\Q_{\rm B}/2$.
The BO is derived from both fRG method and the DW equation method.
The BO transition leads to distinct pseudogap behaviors 
in the NMR $1/T_1$ relaxation rate and in the DOS,
consistently with many experimental reports at $T\approx T^*$.
As we show in the SM B \cite{SM},
very similar numerical results
are obtained in the case of $t'/t=0.7$.
Thus, the $d$-wave BO  
will be ubiquitous in $\kappa$-(BEDT-TTF)$_2$X.
The present theory would be applicable for 
other strongly correlated metals with pseudogap formation.

\acknowledgements
We are grateful to S. Onari for useful discussions.
This work is supported by Grants-in-Aid for Scientific Research (KAKENHI)
Research (No. JP20K22328, No. JP20K03858, No. JP19H05825, No. JP18H01175, JP16K05442)
from MEXT of Japan.

%%%%%%%%%%%%%%%%%%%%%%%%%%%%%%%%%%%%%%%%%%%%%%%%

%\end{document}
%%%%%%%%%%%%%%%%%%%%%%%
\clearpage

%\section{
%[Supplemental Material]
%}

\makeatletter
\renewcommand{\thefigure}{S\arabic{figure}}
\renewcommand{\theequation}{S\arabic{equation}}
\makeatother
\setcounter{figure}{0}
\setcounter{equation}{0}
\setcounter{page}{1}
\setcounter{section}{1}

\begin{widetext}
\begin{center}
{\bf \large 
[Supplementary Material] \\
\vspace{3mm}
{\large
Prediction of $d$-wave bond-order and pseudogap 
in organic superconductor $\kappa$-(BEDT-TTF)$_2$X:
Similarities to cuprate superconductors
%Microscopic theory of bond ordering and superconductivity \\
%in $\kappa$-(BEDT-TTF) based on fRG study
}
}%
\end{center}

\begin{center}
Rina Tazai$^1$, Youichi Yamakawa$^1$, Masahisa Tsuchiizu$^2$ and
Hiroshi Kontani$^1$
\end{center}
\begin{center}
\textit{
$^1$Department of Physics, Nagoya University, Furo-cho, Nagoya 464-8602, Japan. \\
$^2$Department of Physics, Nara Women's University, Nara 630-8506, Japan.
}
\end{center}

\end{widetext}
%%%%%%%%%%%%%%%%%%%%%%%%%%%%%%%%%%%%%%%%%%

%%%%%%%%%%%%%%%%%%%%%%%%%%%%%%%%%%%%%%%%%%%%%%%
\section{A: RG+cRPA method for the dimer Hubbard model
of $\kappa$-(BEDT-TTF)$_2$X
}

\begin{figure}[htb]
\includegraphics[width=.7\linewidth]{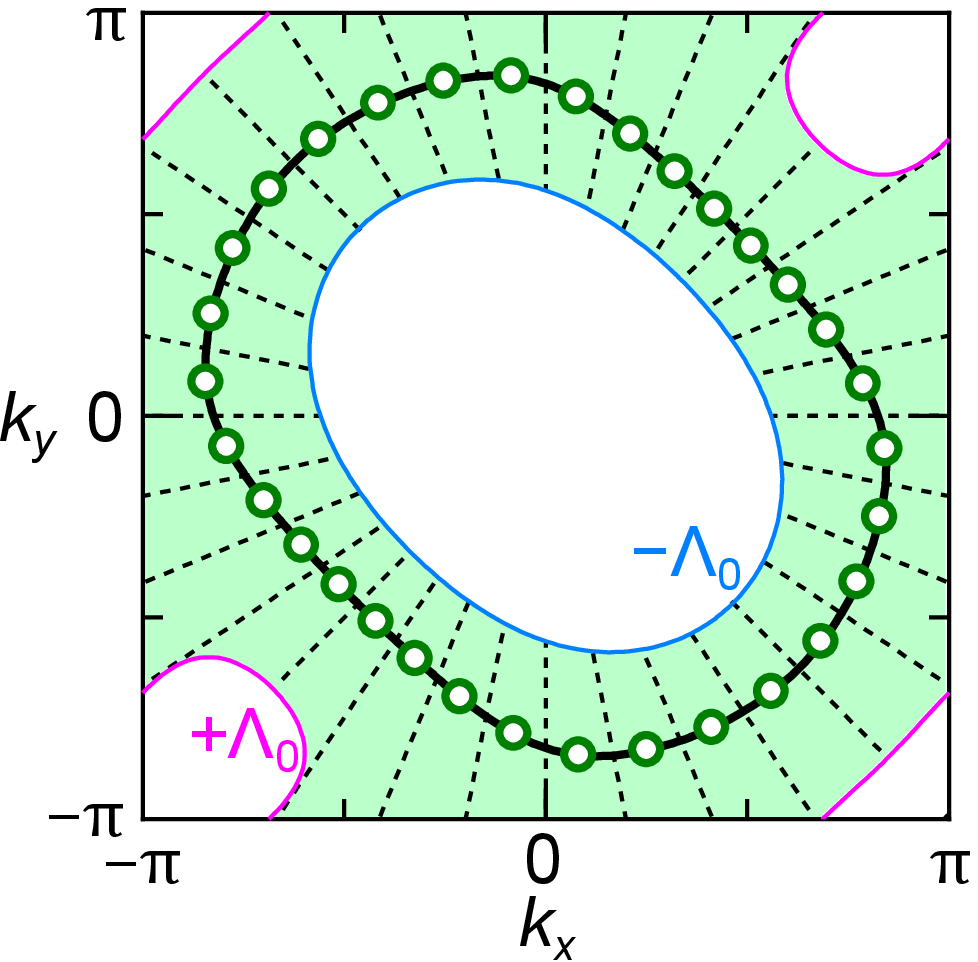}
\caption{
The FS and the BZ of the present dimer model.
The green shaded region $|\e_\k-\mu|\le \Lambda_0=2$
is divided into $N_p\ (=32)$ patches.
The center of each patch on the FS is shown as small circle.
}
\label{fig:figS1}
\end{figure}

In the main text, we studied the dimer Hubbard model,
which is the simplest effective model for $\kappa$-(BEDT-TTF)$_2$X,
by applying the RG+cRPA method.
This is a useful hybrid method
between the fRG theory and the RPA 
developed in Ref. \cite{S-Tsuchiizu-2013}.
The RG+cRPA method is applicable for systems with complex 
valence-band structure, even when the conventional patch RG method
is not applicable.
Here, we solve the one-loop RG equation 
for the four-point vertex $\Gamma$ in Fig. \ref{fig:model} (d)
inside the energy region $|\e_\k-\mu|\le \Lambda_0$,
by applying the logarithmic energy mesh
$\Lambda_{l}=\Lambda_{0}e^{-l}$ with $l\ge0$.
In the framework of the $N_p$-patch RG,
the lower-energy region in the Brillouin zone (BZ) 
is divided into $N_p$ patches, as shown in Fig. \ref{fig:figS1}
with $\Lambda_0 =2$ and $N_p=32$.
(Note that the numerical study in the main text is done for $N_p=64$.)
By solving the RG equation, we can calculate 
the vertex corrections (VCs), which are the 
many-body effects that are dropped in the RPA.

%%%%%%%%%%%%%%%%%%%%%%%%%%%%%%%%
%%%%%%%%%%%%%%%%%%%%%%%%%%%%%%%%

For the higher energy region $|\e_\k-\mu|\le \Lambda_0$,
where the VCs are less important,
we apply the constrained random-phase-approximation (cRPA)
with high numerical accuracy using fine $\k$-meshes.
The obtained effective interaction by cRPA 
is incorporated into the initial parameters of the fRG equation,
without worrying about the double counting of diagrams.
In solving the RG equation,
we calculate the pp and ph scattering processes that include 
(at least) one on-shell state ($\Lambda_l>\e_\k>\Lambda_{l+dl}$)
step-by-step, till the parameter $l$ reaches $\ln(\Lambda_0/\w_c)$.
In the main text, we set $\w_c=\pi T$ in the RG equation for $\Gamma$,
and $\w_c=T/100$ in the RG equations
for the three-point vertex and the susceptibility.
By using the RG+cRPA method,
we can perform reliable numerical calculations
\cite{S-Tsuchiizu-2013,S-Tsuchiizu-CDW1,S-Tsuchiizu-CDW2}.
The obtained numerical results are essentially 
robust against the choice of $\Lambda_0$.

%%%%%%%%%%%%%%%%%%%%%%%%%%%%%%%%%%%%%%%%%%%%%%%%%%%%%
\section{B: 
$d$-wave BO solutions in the dimer Hubbard model with $t'/t=0.7$
}

In the main text, we examined the electronic states in the
dimer Hubbard model with $t'/t=0.5$.
However, the ratio $t'/t$ in $\kappa$-(BEDT-TTF)$_2$X
depends on the anion molecule X.
In many compounds, the relation $t'/t=0.5\sim0.7$ is realized,
except for the spin liquid compound X=Cu$_2$(CN)$_3$
with $t'/t\approx1$.
In our previous study \cite{S-Kino-ET},
experimental AFM-superconducting phase diagram
is reproduced for $t'/t=0.5\sim0.8$, by setting the ratio $U/|t|$
larger for larger $t'/t$.

\begin{figure}[htb]
\includegraphics[width=.99\linewidth]{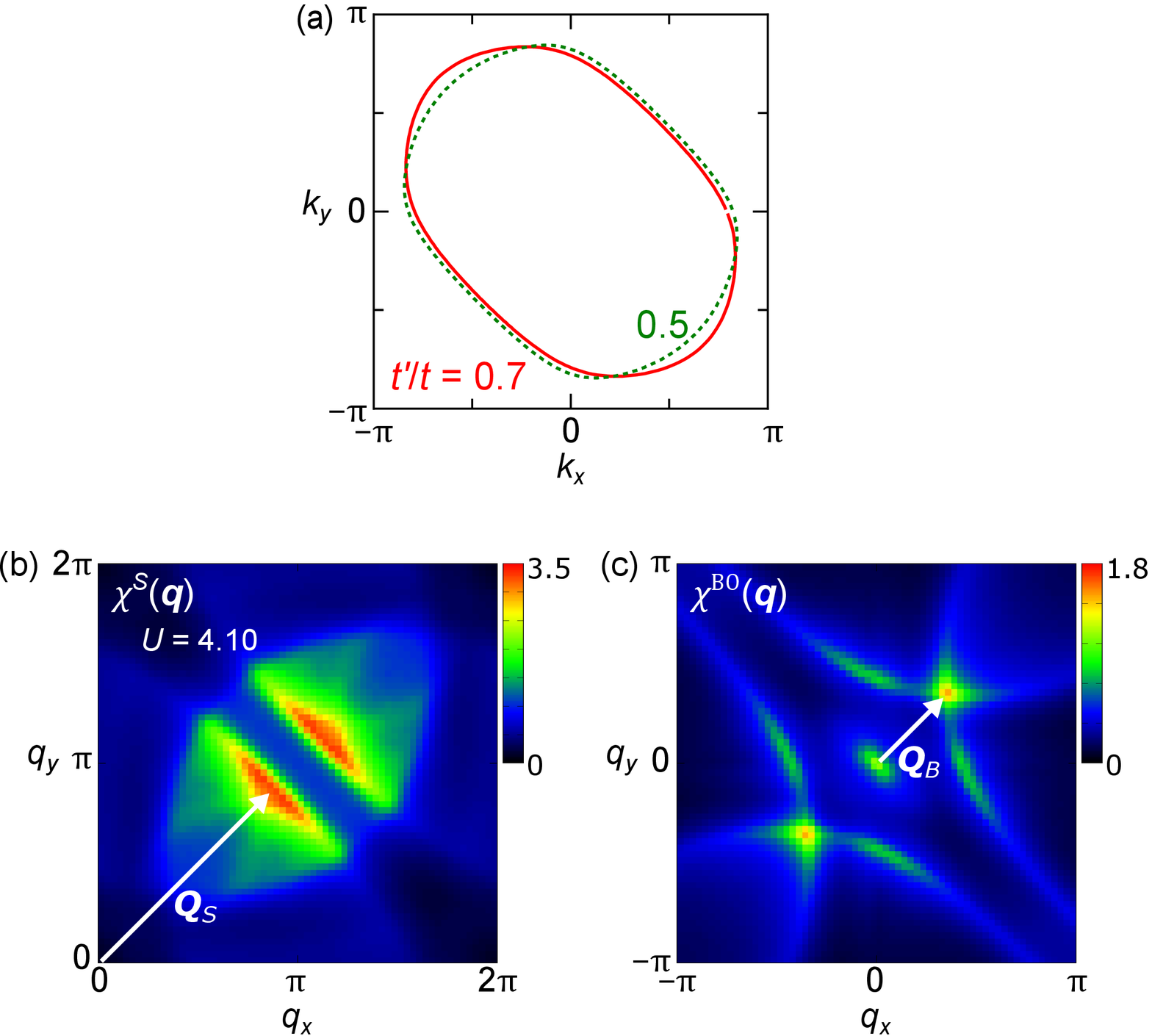}
\caption{
(a) FS of the dimer model for $t'/t=0.7$.
We also show the FS for $t'/t=0.5$, which is used in the main text.
(b) $\chi^{\rm S}(\q)$ and (c) $\chi^{\rm BO}(\q)$ obtained by the 
RG+cRPA method for $U=4.10$ in the case of $t'/t=0.7$.
The obtained $\chi^{\rm BO}(\q)$ has the 
peak structures at both $\q=\bm{0}$ and $\Q_{\rm B}$.
%(d) Obtained RG flow of $\chi^{\rm S}(\Q_S)$, $\chi^{\rm BO}(\Q_{\rm B})$
%and $\chi^{\rm BO}(\bm{0})$ for $U=4.12$.
}
\label{fig:figS2}
\end{figure}

Here, we analyzed the dimer Hubbard model with $t'/t=0.7$
in order to verify the robustness of the $d$-wave BO against the 
modification of the model parameters.
The Fermi surface (FS) for $t'/t=0.7$ is shown in
Fig. \ref{fig:figS2} (a).
The spin susceptibility and $d$-wave BO susceptibility 
obtained by the RG+cRPA method are shown in 
Figs. \ref{fig:figS2} (b) and (c), respectively,
in the case of $U=4.10$ at $T=0.05$.
%The enhancement of $\chi^{\rm BO}(\q)$ originates from the VCs 
Since $\chi^{\rm BO}(\q)$ remains very small in the RPA (FLEX),
the obtained strong BO fluctuations in Fig. \ref{fig:figS2} (c)
originate from the VCs that are dropped in the RPA (FLEX).
The BO susceptibility strongly develops 
with increasing $U$, and it
exceeds the spin susceptibility for $U\ge 4.12$.

%%%%%%%%%%%%%%%%%%%%%%%%%%%%%%%%
%%%%%%%%%%%%%%%%%%%%%%%%%%%%%%%%

%Figure \ref{fig:figS2} (d) shows the obtained
%RG flow of $\chi^S(\Q_S)$, $\chi^{\rm BO}(\Q_{\rm B})$
%and $\chi^{\rm BO}(\bm{0})$ as functions of $l$,
%in the case of $U=4.12$.
%Thus, the $d$-wave BO can be realized above the 
%SDW transition temperature,
%consistently with the numerical results of the main text
%for $t'/t=0.5$.

%%%%%%%%%%%%%%%%%%%%%%%%%%%%%%%%
\begin{figure}[htb]
\includegraphics[width=.99\linewidth]{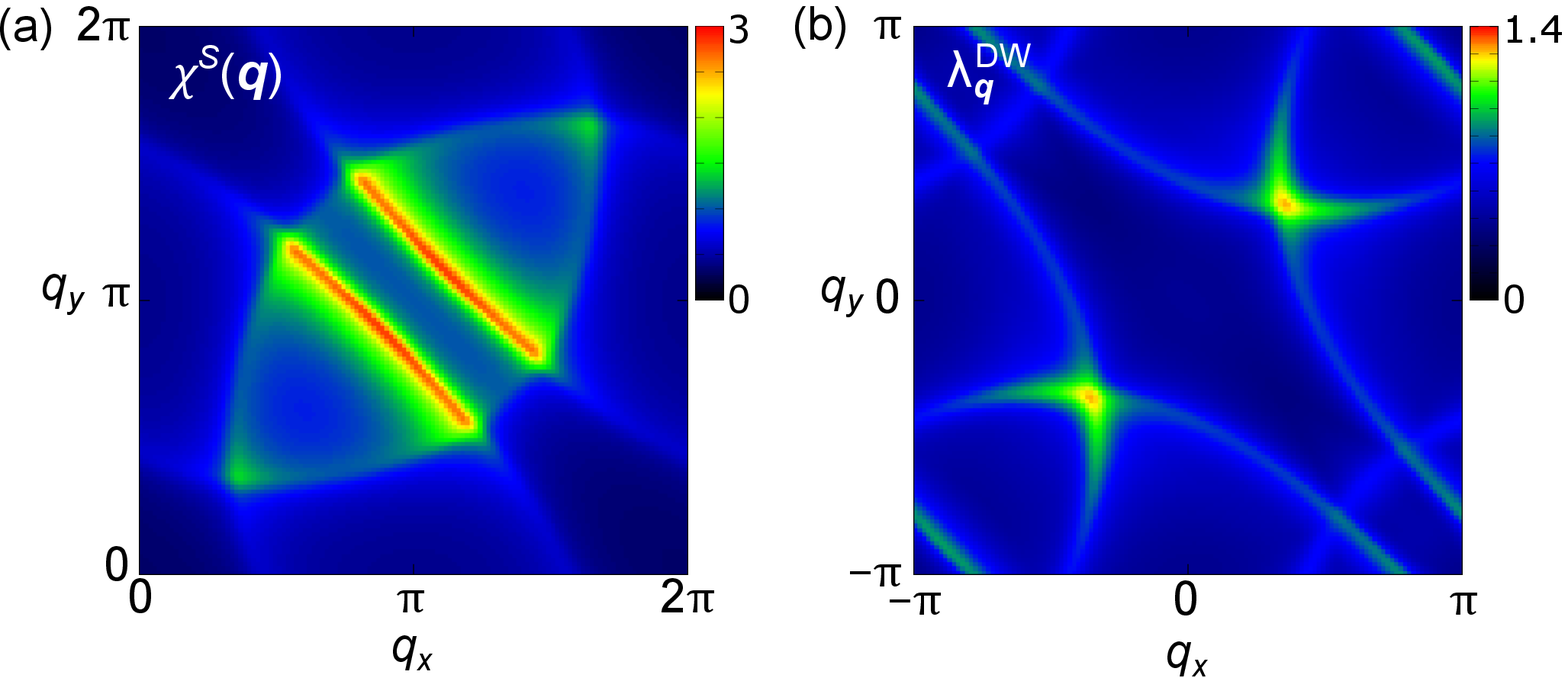}
\caption{
(a) $\chi^{\rm S}(\q)$ and (b) $\lambda^{\rm DW}(\q)$ obtained by the 
DW equation for $U=3.10$ in the case of $t'/t=0.7$.
}
\label{fig:figS3}
\end{figure}
%%%%%%%%%%%%%%%%%%%%%%%%%%%%%%%%

Next, 
we perform the DW equation analysis for $t'/t=0.70$.
Then, the RPA spin susceptibility is shown in 
Fig. \ref{fig:figS3} (a) for $U=3.10$ at $T=0.05$.
where the Stoner factor $\a_S=U\chi^0(\Q_S)$ is $0.90$.
Figure \ref{fig:figS3} (b)
is the obtained eigenvalue of the DW equation.
The derived form factor at $\q=\Q_{\rm B}$ is very similar
to the $d_{x^2-y^2}$-wave form factor for $t'/t=0.5$
obtained in Fig. \ref{fig:fig4} (c) in the main text.
These results are essentially similar to 
the results by the RG+cRPA method in Fig. \ref{fig:figS2},
except for the absence of the peak at $\q=\bm{0}$ 
in Fig. \ref{fig:figS3} (b).

In summary,
the development of the $d$-wave BO susceptibility 
at $\q=\Q_{\rm B}$, $\chi^{\rm BO}(\Q_{\rm B})$,
is confirmed by both the fRG theory and the DW equation 
theory in the dimer Hubbard model,
in the cases of $t'/t=0.5$ and $0.7$.
This result is derived from the AL-type VCs,
which are neglected in the RPA.
Since $\lambda^{\rm DW}_{\q=\bm{0}}$ in Fig. \ref{fig:figS3} (b)
remains small, the enhancement of $\chi^{\rm BO}(\bm{0})$ 
in Fig. \ref{fig:figS3} (c) originates from the 
spin and SC fluctuations cooperatively, 
as we discussed in the main text.
%both $\chi^{\rm BO}(\Q_{\rm B})$ and grow in a parallel way, while

Finally, it should be stressed that the ferro-BO  
can be induced by the paramagnon-interference mechanism 
in general systems.
In fact, strong ferro-BO fluctuations 
observed in both Fe-based and cuprate superconductors
are satisfactorily reproduced by the DW equation analysis
\cite{S-Onari-FeSe,S-Onari-B2g,S-Kawaguchi-CDW},
meaning that the ferro-BO fluctuations originate solely 
from the spin fluctuations.
%It is an important issue to clarify the strength of
%the ferro-BO fluctuations in $\kappa$-(BEDT-TTF)$_2$X.

%In this sense, the reason for the small ferro-BO fluctuations
%in Fig. \ref{fig:figS3} (b) should be verified by future studies.
%(Note that $\lambda_{\q}^{\rm BO}$ shows a peak structure at 
%$\q={\bm 0}$ for $\a_S\sim0.99$.)

%%%%%%%%%%%%%%%%%%%%%%%%%%%%%%%%%%%%%%%%%%%%%%%%

%%%%%%%%%%%%%%%%%%%%%%%

\end{document}